\documentclass[preprint2]{emulateapj}

\slugcomment{accepted for publication in ApJ} \shorttitle{Carbon
nanoparticle Formation} \shortauthors{J\"ager et al.}

\begin{document}
\title{Formation of PAHs and Carbonaceous Solids in Gas-Phase Condensation Experiments}
\author{C. J\"ager\altaffilmark{1}, F. Huisken}
\affil{Max-Planck-Institut f\"ur Astronomie, K\"onigstuhl 17,
D-69117 Heidelberg and Institut f\"ur Festk\"orperphysik,
Helmholtzweg 3, D-07743 Jena, Germany}
\email{Cornelia.Jaeger@uni-jena.de}
\author{H. Mutschke, I. Llamas Jansa}
\affil{Astrophysikalisches Institut und Universit\"ats-Sternwarte
(AIU), Schillerg\"{a}sschen 2-3, D-07745 Jena, Germany}
\author{Th. Henning}
\affil{Max-Planck-Institut f\"ur Astronomie, K\"onigstuhl 17,
D-69117 Heidelberg, Germany}
\altaffiltext{1}{Corresponding author}

\begin{abstract}
Carbonaceous grains represent a major component of cosmic dust. In
order to understand their formation pathways, they have been
prepared in the laboratory by gas-phase condensation reactions such
as laser pyrolysis and laser ablation. Our studies demonstrate that
the temperature in the condensation zone determines the formation
pathway of carbonaceous particles. At temperatures lower than
1700\,K, the condensation by-products are mainly polycyclic aromatic
hydrocarbons (PAHs), that are also the precursors or building blocks
for the condensing soot grains. The low-temperature condensates
contain PAH mixtures that are mainly composed of volatile 3-5 ring
systems. At condensation temperatures higher than 3500\,K,
fullerene-like carbon grains and fullerene compounds are formed.
Fullerene fragments or complete fullerenes equip the nucleating
particles. Fullerenes can be identified as soluble components.
Consequently, condensation products in cool and hot astrophysical
environments such as cool and hot AGB stars or Wolf Rayet stars
should be different and should have distinct spectral properties.
\end{abstract}


\section{INTRODUCTION}
It is generally assumed that most of the primary cosmic carbonaceous material is formed
as nano- and subnanometer-sized particles via gas-phase condensation in envelopes
of carbon-rich asymptotic giant branch (AGB) stars \citep{Henning:98}. The very probable and abundant precursor for this
process is acetylene (C$_{2}$H$_{2}$). The presence of a solid nanosized
carbonaceous material in addition to molecular polycyclic aromatic hydrocarbons (PAHs), which are considered
to be the carriers of the aromatic IR (AIR) bands in different astrophysical environments,
is sufficiently proven by the IR excess emission of transiently heated grains \citep{Boulanger:98,Draine:01}.
However, a firm detection of individual band
carriers for the AIRs has not been succeeded. The role of fullerenes as possible band
carriers remains uncertain. Giant fullerenes were already found in
primitive meteorites \citep{Smith:81,Buseck:02}. The detection of
fullerenes in astronomical environments is still a matter of debate
\citep{Foing:97,Galazutdinov:00,Nuccitelli:05,Sellgren:07}.

Several authors \citep{Frenkl:89,Cherchneff:92, Allain:97} have modeled the
formation of PAHs in AGB stars. Whereas these authors determined a
narrow temperature range of 900 to 1100\,K for the formation of PAHs in circumstellar
environments, \citet{Cherchneff:99}
reconsidered the modeling of the PAH and carbon dust formation in
carbon-rich AGB stars. They developed a physico-chemical model,
which describes the periodically shocked gas in the circumstellar
shells close to the photosphere of the stars. The authors found that
benzene (C$_6$H$_6$) formation begins at 1.4 R$_*$ and, at a radius of 1.7
R$_*$, the conversion of single rings to PAHs starts at a temperature of around 1700\,K.
This temperature is much higher than the temperature window calculated in previous studies.

Temperatures and pressures in circumstellar condensation environments are uncertain
since they depend on various model assumptions, for instance, mass loss rates, gas density
outflow velocity, and dust formation. Pulsations of the helium burning shells
affect the atmospheres and can cause shock fronts, which result in density/pressure
fluctuations \citep{Nowotny:05}. Pressure ranges of 0.03--100 and 1--800\,dyn/cm$^{2}$
(3$\times$10$^{-5}$--0.1 and 1$\times$10$^{-3}$--0.8\,mbar) for AGB stars with an
effective temperature of 2500 and 4000\,K, respectively, have been calculated \citep{Lederer:06}.

One important way to successfully understand the condensation of
dust in astrophysical environments is to calculate the stability
limits for condensation of dust species based on chemical models and
different chemical and physical parameters. \citet{Lodders:99} have
modeled the condensation of carbonaceous grains in dust-forming
shells of carbon-rich AGB stars using pressures between 10 and
3$\times$10$^3$\,~dynes/cm$^2$ (0.01-3\,~mbar). The authors found
that the condensation temperatures generally decrease with
decreasing total pressure. However, graphite is an exception since
the condensation temperature is not pressure-dependent but sensitive
to the C/O ratio in the dust-forming zone. Higher C/O ratios shift
the pressure range to higher values.

For hydrogen-poor Wolf Rayet (WR) stars with high effective temperatures between 20,000 and
90,000\,K, running through the WC phase, the carbon dust condensation
is supposed to start with small carbon chains, that can form monocyclic
rings \citep{Helden:93}. These carbon chains and rings represent the precursors for
fullerenes, which finally form the amorphous carbon grains \citep{Cherchneff:00}.
In this scenario, the PAHs are ruled out as possible intermediates.

It is imperatively assumed that the formation pathway of carbon condensation determines the
resulting structure, chemical composition, and morphology of the condensing
grains. Therefore, it is important to know the formation process of cosmic dust grains
and possible precursors and/or intermediates such as PAHs, fullerenes, or polyyne
carbon chains, in order to understand their spectral properties in different astrophysical environments.

Despite all efforts, the formation process of carbon nanoparticles in astrophysical environments
and even in terrestrial carbon condensation processes is not
sufficiently understood. Most studies of soot formation pathways have been performed
on premixed gas flames or laminar diffusion flames.
The formation of PAHs and nanometer-sized carbon grains in fuel-rich hydrocarbon flames has been
extensively investigated by \citet{Burtscher:92}, \citet{Baum:92}, and \citet{Weilmuenster:99}.
The chemical species associated with the early stage of soot growth and soot precursor particles
in flames have been studied by \citet{Dobbins:98} and \citet{Oektem:05}.
\citet{Homann:98} claimed that, by association of two or more large PAHs in form of biaryl or $\pi$
complexes, a new group of PAHs is formed, which he called aromers. These aromers
are supposed to be precursors for fullerenes and soot particles.
The disadvantage of studying soot formation in flames is the varying temperature and
chemistry in the flames that seems to be associated with changes in the particle formation pathways.

We have employed gas-phase condensation experiments to produce
nanometer- and subnanometer-sized carbonaceous particles within two
different temperature regimes, a high-temperature (HT) and
low-temperature (LT) condensation regime. Both gas-phase
condensation techniques come close to the grain formation process
suggested to occur in astrophysical environments. The study of the
formation pathways has been performed experimentally by analytical
characterization of the soot and its by-products, including
high-resolution transmission electron microscopy (HRTEM),
chromatographic methods, and mass spectroscopy. Finally, the
spectral properties of the two different condensates have been
measured from the far UV (FUV) to the infrared (IR), in order to
compare their spectral characteristics. They are are discussed in
$\S$ \ref{spec}.

\section{EXPERIMENTAL CONDITIONS FOR HIGH- AND LOW-TEMPERATURE CONDENSATION}
\label{experi}
The synthesis conditions for all series of gas-phase condensation
experiments are presented in Table\,\ref{prod}. The first two rows
refer to the high-temperature (HT) condensation experiments performed with pulsed lasers characterized by
high power densities and, consequently, high temperatures of more
than 3500\,K in the condensation zone. The last four rows of Table\,\ref{prod} describe
the low-temperature (LT) condensation experiments that process at temperatures lower than 1700\,K.
The last column of the table contains a description of the two types of condensed carbonaceous matter.

For HT condensations, laser ablation of graphite with subsequent
condensation of carbonaceous matter in quenching gas atmospheres of
He or He/H$_2$ (3/2 He/H$_2$ standard volume flow) mixtures at
pressures between 3.3 and 26.7\,mbar has been applied. The second
harmonic of a pulsed Nd:YAG laser was used to evaporate carbon from
a graphite target. The pressure was kept as low as possible to come
close to the conditions in astrophysical condensation zones.
However, at a pressure lower than 3.3\,mbar, the condensation rate
became too slow to be manageable for the condensation of grains. In
order to study the influence of the pressure on the composition and
final structure of the condensate, we have increased the pressure up
to 26.7\,mbar. In addition, laser-induced pyrolysis (LIP) of
gas-phase hydrocarbon precursors (ethylene, acetylene, and benzene)
at pressures around 335\,mbar using a pulsed CO$_2$ laser have been
carried out. In such a condensation experiment, the pressure has to
be higher in order to obtain a stable flame and constant
condensation conditions. In the table, the terms LA (laser ablation)
and LPPL (laser pyrolysis, pulsed laser) refer to a number of
different experiments with varying laser power densities, quenching
gas mixtures, pressures, and precursors. The condensing particles
were extracted from the condensation zone by using a molecular beam
technique and deposited on KBr and CaF$_2$ substrates for IR and
UV/visible spectroscopy, respectively. A more detailed description
of both methods can be found in \citet{Llamas:07} and
\citet{Jaeger:08b}. For all the HT experiments, the laser power
densities were varied between 1$\times$10$^7$ and
9$\times$10$^9$\,W\,cm$^{-2}$, which is a measure that correlates
with the temperature in the condensation zone. For HT gas-phase
condensation processes of nanoparticles, we can provide lower limits
of the temperature in the condensation zone. The vibrational
temperature of the laser-induced plasma generated by laser
evaporation of a graphite target in a 10\,mbar He atmosphere was
found to range between 4000 and 6000\,~K for power densities between
5$\times$10$^8$--2$\times$10$^9$\,W\,cm$^{-2}$ \citep{Iida:94}. In
the LIP of hydrocarbons, the employed power densities are only
slightly lower compared to the ablation experiments and, therefore,
the condensation temperatures are well comparable, which is
confirmed by other authors \citep{Koji:81,Doub:06}.

LT condensation studies have been performed by laser-induced
pyrolysis (LIP) of gas-phase hydrocarbon precursors, such as
ethylene, acetylene, and benzene, at pressures around 750\,mbar,
applying a continuous-wave (cw) CO$_2$ laser. The detailed
experimental setup is described elsewhere
\citep{Jaeger:06a,Jaeger:06b}. The condensed nanoparticles have been
collected in a filter. The terms LPcw2--LPcw5 refer to individual
condensation experiments performed by application of laser pyrolysis
with a cw CO$_2$ laser. For the LT condensation experiments, the
laser power densities varied between 850 and 6400\,W\,cm$^{-2}$. The
power densities are orders of magnitude lower compared to the HT
condensations and, therefore, the corresponding temperatures in the
condensation zone are much lower and were found to be in the range
between 1000 and 1700\,K. In these experiments, the temperatures
have been determined with a pyrometer.

For both, HT and LT condensations, the pressures considerably
exceeded the pressure ranges supposed to be valid in the
condensation zone of AGB stars. In laboratory experiments it is
necessary to scale up the pressure since the free path lengths of
atoms and clusters would be too long for the laboratory setup at
pressures lower than 1\,mbar. Therefore, in order to simulate the
condensation of particles, we have to scale up the applied pressure
ranges in the condensation apparatus at least by 1-2 orders of
magnitude. However, we would like to point out that the total
pressures provided in Table\,\ref{prod} do not represent the partial
pressures of the condensing species since we additionally use a
quenching gas such as helium or hydrogen. The real partial pressure
of the condensing carbon species cannot be measured. Therefore, an
exact determination of the ratio between the partial and the
equivalent pressure of the gaseous carbon species, which is related
to the supersaturation factor (see $\S$ \ref{result}), during the
condensation process is very difficult.

The internal structures of the condensed carbon particles and the formation of possible
intermediates and by-products were investigated with high-resolution
transmission electron microscopy (HRTEM) (JEOL JEM 3010 microscope)
operating at an acceleration voltage of 300\,kV. The soluble components of the condensates were removed by
soxhlet extraction in toluene. The composition of the extracts and
condensates was characterized by using gas chromatographic/mass spectrometric (GC/MS) analyses,
high-performance liquid chromatography (HPLC) combined with a UV/visible
diode array detector, and matrix-assisted laser desorption/ionization
in combination with time-of-flight mass spectrometry (MALDI-TOF).
Technical details of the analytical and spectral analyses are provided in previous papers \citep{Jaeger:06a,Jaeger:08b}.
\begin{table*} \caption[]{Experimental conditions comprising the HT and LT gas-phase
condensation experiments. The terms LA and LPPL denote a series of experiments performed at
high temperatures, but with varying conditions. The terms LPcwi describe individual LT condensation experiments.
A more detailed description of these experiments can be found elsewhere
\citep{Llamas:07,Jaeger:08b}.} \label{prod}
\begin{center}{\scriptsize
\begin{tabular}{ccccccc}
\hline
Experi- &Precursor             &  Buffer & Laser  & Laser power                     & Tempera-     &  Condensate              \\
ment    &                      &  gas    &        &  density (W\,cm$^{-2}$)         & ture (K)     &                     \\
  \hline
LA      &Graphite              & He/H$_2$& pulsed &2$\times$10$^8$--9$\times$10$^9$ & $\geq$4000  &  fullerene-like soot\\
        &                      &         &        &                                 &             &  and fullerenes     \\[0.1cm]
LPPL    &C$_2$H$_4$, C$_2$H$_2$,& He/Ar   & pulsed &1$\times$10$^7$--1$\times$10$^9$ & $\geq$3500  &  fullerene-like soot\\
        &C$_6$H$_6$            &         &        &                                 &             &  and fullerenes     \\
\hline
LPcw2     &C$_2$H$_4$, C$_2$H$_2$& Ar      &  cw    & 5200                            &             &soot and 14 wt\% PAHs\\[0.1cm]
LPcw3     &C$_2$H$_4$, C$_6$H$_6$& Ar      &  cw    & 5200                            & $\sim$1500  &soot and 33 wt\% PAHs\\[0.1cm]
LPcw4     &C$_2$H$_4$, C$_6$H$_6$& Ar      &  cw    &  6400                           & $\sim$1700  &soot and 17 wt\% PAHs\\[0.1cm]
LPcw5     &C$_2$H$_4$            & Ar      &  cw    &  850                            & $\sim$1000  &  100 wt\% PAHs      \\
\hline
\end{tabular}}
\end{center}
\vspace{1mm}
\end{table*}

\section{RESULTS AND DISCUSSIONS}\label{result}
\subsection{Soot Formation at Different Temperatures}\label{struc}
\begin{figure}[t]
\epsscale{1.2} \plotone{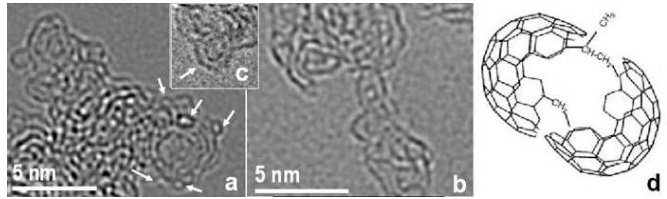}
\caption{HRTEM images of fullerene-like nanoparticles generated in a HT condensation process
 (samples from series LA). In image (a), fullerene and elongated fullerene molecules
 are marked by arrows. Image (b) and (c) present a typical fullerene
 cage fragment and a buckyonion with two interleaved fullerene cages
 (size 1.7 nm, also marked by an arrow), respectively. Model (d)
 illustrates the likely link between fullerene fragments in the soot grains.} \label{HRTEM 1}
\end{figure}

HRTEM images of the carbonaceous matter produced in a HT gas-phase
condensation process are shown in Figure\,\ref{HRTEM 1}.
A much more detailed chemical and structural description of the HT
condensates can be found in \citet{Llamas:07} and
\citet{Jaeger:08b}. The HRTEM micrographs reveal that, in addition
to individual fullerenes, which are marked by arrows in
Figure\,\ref{HRTEM 1}a, very small fullerene-like particles are
produced. These particles are composed of small, strongly bent
graphene layers with varying lengths and distances between these
layers. The level of disorder depends on the employed condensation
conditions. Soot condensates containing low amounts of hydrogen show
more ordered and frequently closed fullerene cages
(Figure\,\ref{HRTEM 1}c) whereas grains with higher hydrogen
contents are less ordered and do not show completely closed cages.
In these grains, the cage fragments stick together by van der Waals
forces or are linked by aliphatic --CH$_x$ groups (see sketch d in
Figure\,\ref{HRTEM 1}).

Both methods of HT condensation (laser ablation and laser pyrolysis)
work at different pressures. However, the increase of the pressure
in the condensation zone by two orders of magnitude does not result
in a change of the structure of the condensate.

The small size of the grains points to a strong supersaturation of
the carbon vapor in the condensation zone resulting in a high number
of nucleation seeds. This is based on the assumption that,
generally, homogeneous nucleation needs a high supersaturation of
the vapor phase \citep{Granquist:76,Heidenreich:03}. The relation
between the size of the critical grains $r^*$ and the Gibbs free
energy is defined by the equation $r^* = -2\gamma / \Delta G_v$,
where $\gamma$ stands for the surface energy of the condensed
species \citep{Cao:04}. The Gibbs free energy $\Delta$$G_v$  is
again proportional to the temperature $T$ and the supersaturation
$S$ via the relation $\Delta G_v = \frac{-kT}{\Omega} ln(1+S)$ where
$k$ is the Boltzmann constant and $\Omega$ is the atomic volume.
Therefore, in order to reduce the critical size of stable nuclei in
the condensation zone, which is a requirement for the production of
small grains, one needs to increase the Gibbs free energy
$\Delta$$G_v$ of nucleation per unit volume. TEM micrographs of the
HT samples partly show the appearance of larger particles up to
10\,nm, which were formed by coagulation of the original, very small
fullerene-like carbon grains. Under these condensation conditions
(high supersaturation and temperature), it is supposed that the
further particle growth is exclusively due to coagulation.

\begin{figure}[htp]
\epsscale{0.8} \plotone{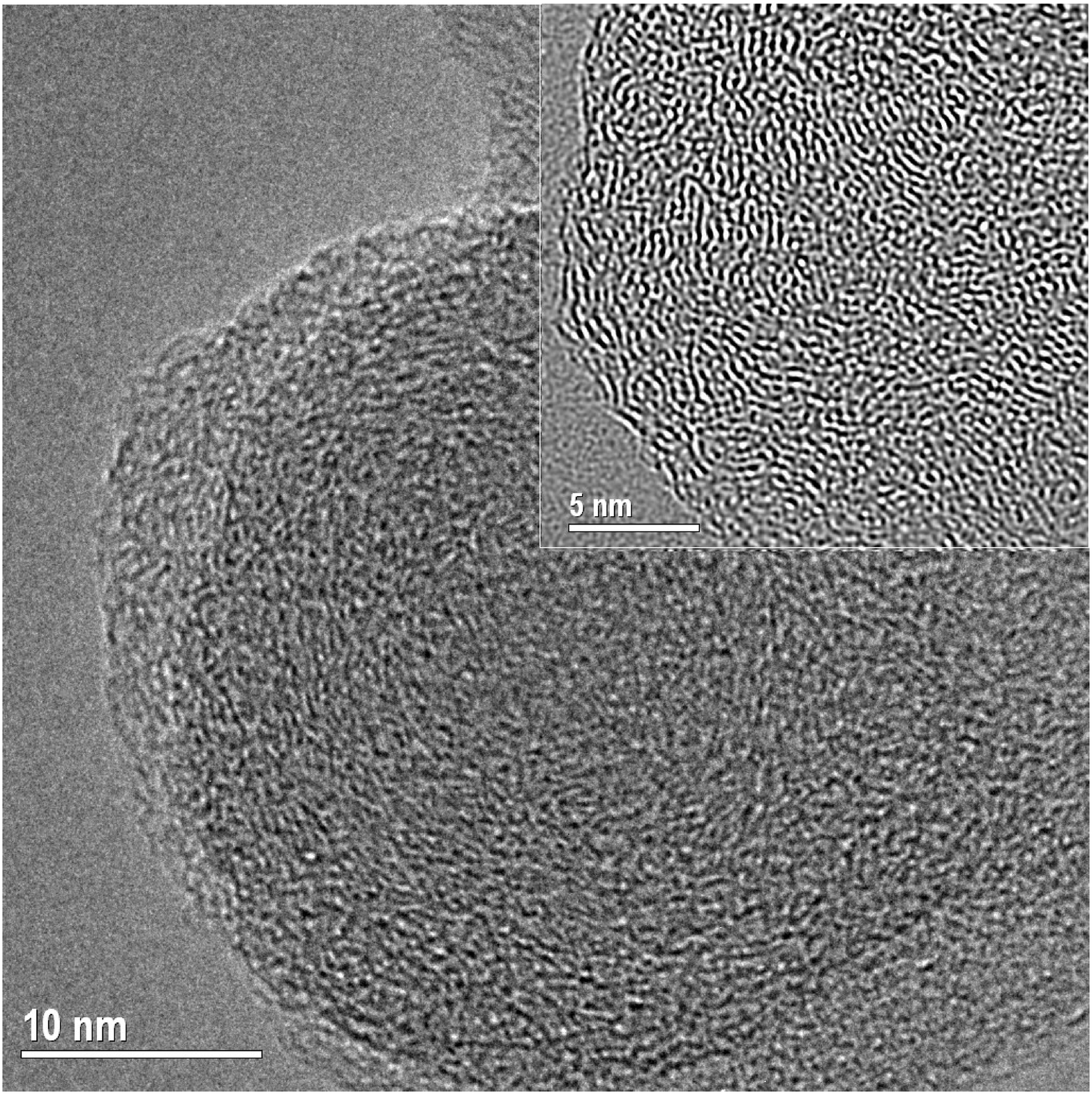} \caption{HRTEM image of a typical
large carbon nanoparticle produced in a LT process (LPcw2-4). The
inset displays a section of a particle showing the arrangement and
mean sizes of the graphene layers.} \label{HRTEM 2}
\end{figure}
The generation of fullerenes and fragments of them in the condensation process
could be verified by using electron microscopy.
The presence of symmetric and elongated fullerene molecules with different numbers of carbon atoms
is clearly visible in the HRTEM micrographs. The symmetric fullerenes range from 0.5 up to 1.03\,nm
corresponding to fullerenes from C$_{36}$  up to C$_{130}$ \citep{Goel:04}.
A soot formation process via polyyne chains, fullerenes and fullerene snatches,
respectively, was already proposed by \cite{Kroto:88}.
MALDI-TOF studies of the HT condensate have shown that no PAHs are formed
as intermediates.

The critical point in the creation of bent graphene layers or disturbed fullerene-like
structures is the formation of cage fragments or molecules from the chains.
Quantum chemical molecular dynamics simulations have demonstrated that
high carbon densities are essential for such processes, which can be found
for example in carbon arc or laser ablation processes \citep{Zheng:05}.
These authors found that the C$_2$
molecules quickly combine to long and branched carbon chains and
macromolecules for temperatures above 2000\,K, which form small cyclic
structures with long carbon chains attached (nucleation).
The second step is a further ring condensation
growth for example between two linear chains attached to a nucleus.
Consequently, fullerene fragments of bowl shape with side chains are
formed in this step \citep{Irle:03}. Furthermore,
coalescence reactions of small fullerenes \citep{Yeretzian:93}
can result in the formation of larger and elongated fullerene cages,
which have also been observed by HRTEM.

\begin{figure}[t]
\epsscale{1.05} \plotone{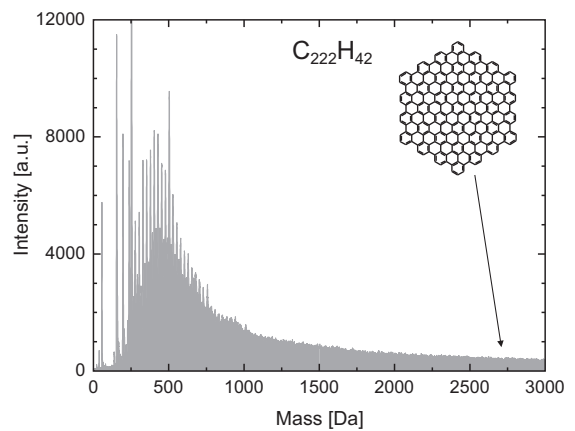} \caption{MALDI TOF spectrum of the
LT soot sample LPcw3 showing that PAHs with masses up to 3000 Da are
observed. The individual peaks show abundant PAHs of lower masses.
To exemplify a PAH molecule with comparable mass, a symmetric
molecule comprising 91 condensed rings and having a diameter of
about 3\,nm is presented.} \label{Maldi}
\end{figure}

In contrast, the condensed grains in LT processes are much larger
and show more ordered and well developed planar graphene layers
inside the particles (see Figure\,\ref{HRTEM 2}). Additionally,
mixtures of PAHs were formed as by-products. These soluble components
were extracted from the soot, and the constituents
of the extract were identified by chromatographic
methods. A mixture of around 70 different PAHs, partly hydrogenated at the edges, could
be identified using GC/MS and HPLC. PAHs with 3-5 ring systems are by far
the most abundant components. PAHs with masses up to
3000 Da were also detected by application of MALDI-TOF mass spectrometry (see Figure\,\ref{Maldi}), but only in
very small amounts. The large size and high internal order of the
condensed soot grains points to a low supersaturation of the carbon vapor
and the formation of a smaller number of stable nuclei compared to the HT condensation
process. In the LT condensation, the further particle growth is
dominated by condensation of intermediates on the surface of the
seeds, which means that PAHs continuously accumulate on the surface
of the grains.
Since larger PAHs have a lower volatility compared to
the smaller ones, a preferred accumulation of large molecules during
the surface growth process can be observed. The accumulation of the
large molecules on the surfaces of the seeds can be confirmed by
a detailed analysis of the HRTEM micrographs revealing a mean
length of the graphene sheets inside the particles of around
1.8\,nm corresponding to PAHs with masses of
around 1000\,Da. However, the longest graphene layers that could be
determined in the micrographs of the soot particles had extensions of
about 3 nm, which corresponds to the highest-mass PAHs detected with
MALDI-TOF measurements.

\subsection{Spectroscopic Properties of the LT and HT Condensates}\label{spec}
\begin{figure}[htp]
\epsscale{1.10} \plotone{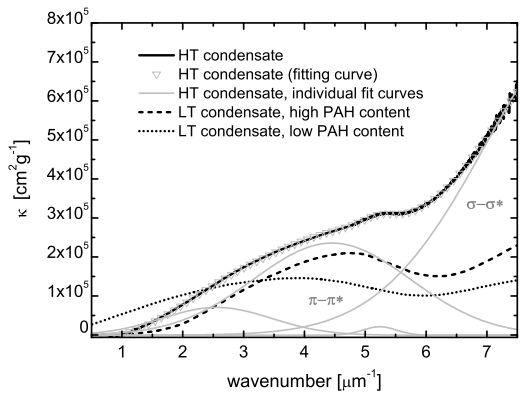} \caption{UV/visible spectra of
selected LT and HT condensate samples. The HT soot sample (black
curve) has been produced in 6.0\,mbar He/H$_2$ atmosphere by laser
ablation (H/C= 0.52). The solid gray curves represent 4 individual
Gaussian profiles fitting as a sum the absorption profile. The
dashed and dotted black curves show the UV/visible spectra of two LT
condensates containing a high and low amount of the soluble organic
component, respectively (see text).}\label{UV}
\end{figure}

The spectral properties of the HT and LT condensates in the
UV/visible range differ considerably. Although the HT soot grains
consist of about 50\% $sp^2$ hybridized carbon atoms, the UV spectra
do not show distinct UV absorption bands (see Figure\,\ref{UV}).
However, a deconvolution of the UV/visible absorption profile by
employing four Gaussians reveals weak and broad bands due to
($\pi$--$\pi$*) transitions between 4.45 and 3.84\,$\mu$m$^{-1}$
depending on the hydrogen content. The width of the ($\pi$--$\pi$*)
main band and the appearance of a plasmon band around
2.5\,$\mu$m$^{-1}$ (400\,nm) accounts for a certain disorder in the
carbon structures due to a broad distribution of curvatures and
lengths of graphene layers in the small fullerene-like carbon grains
\citep{Jaeger:08b,Jaeger:08}. A very weak band around
5.2\,$\mu$m$^{-1}$ (190\,nm) can be attributed to the absorption of
--C=O groups.

In contrast to the HT grains, the soot grains in the LT condensate have much higher content of $sp^2$
hybridized carbon atoms which was found to range between 80 and 92\,\% \citep{NEXAFS:09}.
The UV spectra of LT condensates with very similar grain structure and composition
(soot and PAH mixtures) produced by propane combustion \citep{Schnai:05,Llamas:08}
are shown in Figure\,\ref{UV}. The condensates do show distinct UV absorption bands
caused by ($\pi$--$\pi$*) transitions ranging between 3.6 and 4.9\,$\mu$m$^{-1}$
(204 and 256\,nm). The exact position of this band depends on the internal structure
of the soot grains and the content and composition of the soluble PAH component, which also
contributes to the spectrum. The spectrum with the band position at 3.6\,$\mu$m$^{-1}$
belongs to a condensate consisting of more than 90\,\% soot. Here the band position
is dominated by the soot fraction and shows a strongly enhanced absorption in the visible range.
The second LT spectrum is typical for a composite consisting of
a high amount of PAHs (around 50\,\%) and a smaller fraction of soot grains. The sizes and the internal structure
of the grains are strongly comparable in both condensates. Therefore, the difference in the
positions of the ($\pi$--$\pi$*) transitions is caused by the varying PAH content.

Spectral observations of carbon-rich late-type stars in the UV
range, the places where grain condensation is expected, are limited
since such stars are too cool to be observable in this range. Post
AGB stars and planetary nebulae are brighter in the UV, but due to
the strong rise of temperature, the originally produced dust has
already been processed \citep{Kwok:99}.

The UV spectrum recorded for the active mass-losing and
dust-producing post AGB star HD 44179, recently presented by
\citet{Vijh:05a}, shows a very broad hump with a maximum near
200\,nm (5\,$\mu$m$^{-1}$) which is in width and position similar to
the UV spectrum for the LT condensate containing a high amount of
PAHs (dashed curve in Figure\,\ref{UV}). The effective temperature
of the star amounted to 8000\,K, but the measured blue luminescence
points to the presence of small PAHs. Possibly similar to the
measured UV spectrum in HD 44179 is the one in the hydrogen-rich
post AGB star HD 89353 \citep{Buss:89} with an effective temperature
of 7500\,K \citep{Monier:99}. Aromatic IR bands, observable in both
objects, suggest the presence of PAHs or aromatic subunits in the
carbonaceous dust which probably result from the processing of the
original dust that is supposed to start with the end of the AGB
phase \citep{Kwok:99}.

In the hydrogen-poor post AGB star HD 213985 \citep{Buss:89} and in
the R Coronae Borealis variable stars (R\,CrB and V\,348\,Sgr)
\citep{Hecht:84,Drilling:89,Drilling:97}, UV band positions at 230
(4.34\,$\mu$m$^{-1}$) and 250\,nm (4.0\,$\mu$m$^{-1}$) were
observed, respectively, that are comparable to UV spectral bands
measured for LT condensates but with low PAH content. Indeed, these
objects do not show aromatic IR bands.

The IR spectral properties of the HT condensates are characterized
by strong saturated and aliphatic --CH$_x$ absorptions at 3.4, 6.8,
and 7.25\,$\mu$m. This supports the assumption that saturated
aliphatic --CH$_x$ groups are mainly responsible for the links
between the fullerene fragments (compare Figure\,\ref{HRTEM 1}d). No
aromatic =C--H stretching vibrational bands at 3.3\,$\mu$m have been
observed, but aromatic --C=C-- groups can be identified between 6.2
and 6.25\,$\mu$m. Additionally, out-of-plane bending vibrations of
aromatic =C--H groups are observed between 11 and 14\,$\mu$m, since
their intensity is higher than the intensity of the stretching
bands. A weak feature at 5.8\,$\mu$m can be attributed to a small
amount of --C=O groups inside the carbon structure, but also higher
fullerenes can show vibrational modes in this range
\citep{Mordkovich:00}. Interestingly, signatures for the presence of
--C$\equiv$C-- triple bonds can be observed in the in situ IR
spectra of the HT condensates at 3.03 and 4.7\,$\mu$m, which points
to the formation of polyynes as intermediates and supports the idea
about the formation of fullerene snatches. A more detailed
description of the IR spectral properties of these particles
produced in HT condensations can be found in \citet{Jaeger:08b}.

\begin{figure}[htp]
\epsscale{1.00} \plotone{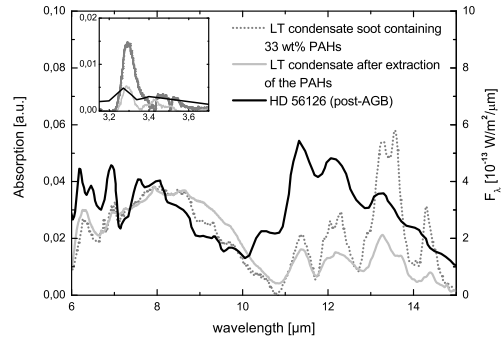} \caption{Comparison between the IR
spectral properties of a LT condensate (LPcw3) and the observed
spectrum of a post AGB star adapted from Hony et al.
(2003).}\label{IR}
\end{figure}

LT condensates, that represent a mixture of soot and PAHs, show
aromatic IR bands (AIBs) as well as aliphatic IR band, with both
types being partly superimposed on two broad plateau features around
8 and 12\,$\mu$m (see Figure\,\ref{IR}). In the spectra, quite a
number of partly small, individual bands can be recognized. However,
a firm separation of aliphatic and aromatic IR features in the
6--10\,$\mu$m range and an attribution of every IR band to a special
functional group of specific components in this range is difficult
and suffers from the fact that the soluble component of the
condensate contains around 70 different aromatic species which can
be partly hydrogenated. In addition, the structural variety of the
grains and their functional groups becomes manifest in additional IR
bands.

The non-treated condensate (see dashed curve) shows aromatic bands
at 3.3, 6.27, 6.35, 6.88, 7.22, 7.93, 8.47, 8.65, 9.28, 9.72, 10.50,
11.36, 11.9, 12.05, 12.14, 12.30, 13.27, 13.56, and 14.30\,$\mu$m as
well as aliphatic IR bands at 3.4, 6.75, 6.94, 7.01, and
7.5\,$\mu$m. The bands beyond 6.7\,$\mu$m mainly arise from --C--C--
stretching and C-H deformation bands. In this spectral range a firm
assignment of an IR band to a specific functional group is very
difficult even for carbonaceous grains without adsorbed PAHs which
may contain a lot of different functional groups with single and
double bonds integrated in varying chemical environments leading to
a distribution of vibrational bands in this spectral range. IR bands
at wavelengths larger than 11\,$\mu$m can be assigned to aromatic
=C--H out-of-plane bending vibrations. In particular, the bands at
13.27, 13.56, and 14.3\,$\mu$m are caused by 4-5 adjacent H atoms
bound to the aromatic ring and argue for smaller PAHs, PAHs with
outer rings containing 4 hydrogen, or for special PAHs containing
phenyl rings bound to larger PAH molecules. In the spectrum of the
soot without PAHs (after toluene extraction), a few small and
distinct bands disappear and the overall spectrum becomes smoother
(see gray solid curve). However, most of the features persist
pointing to the fact that carbonaceous grains can show similar bands
as PAHs. However, very large insoluble PAHs may also remain and
contribute to the IR bands after extraction.

The spectral characteristics of LT condensates resemble the observed
IR spectra of post AGB stars and protoplanetary nebulae
\citep{Kwok:01,Hony:03,Hrivnak:07}. The comparison shown in
Figure\,\ref{IR} clearly reveals that nearly all of the observed IR
bands are also present in the LT condensate produced in the
laboratory. Differences in band ratios, apparent for the =C-H
out-of-plane vibrational bands in the range between 11 and
14\,$\mu$m for the condensate containing soot and PAHs, result from
the overabundance of small PAHs, or larger PAHs containing outer
rings with up to 4 to 5 H atoms compared to larger, compactly
condensed species which should be more stable under astrophysical
conditions. The pure soot particles (after extraction) which have
consumed all the large PAHs during the growth process show a much
better coincidence with the observed bands (see gray solid curve in
Figure\,\ref{IR}). The comparison reveals that the LT condensate is
a promising dust analog for carbonaceous materials produced in
carbon-rich AGB stars.

\section{CONCLUSIONS}
The chemical pathways for the formation of carbon nanoparticles in
HT and LT condensation processes are completely different, resulting
in different structures and spectral properties of the condensed
carbonaceous matter.

The soot grains formed in HT condensation processes are characterized by very small particles
with sizes up to 3-4\,nm. Their structure can be
described as fullerene-like soot grains containing elongated or
symmetric cages, which can be interleaved. Most of the observed structural units
are cage fragments which can be linked to another fragment either by aliphatic
bridges or by van der Waals forces. The formation pathway of the soot particles is characterized
by the formation of carbon chains that can form fullerene fragments of bowl shape
with side chains. Long and branched carbon macromolecules are suggested to be
the precursors for cyclic structures with long carbon chains attached which can condense
and grow, for example, by a linkage of two linear chains attached to a nucleus \citep{Irle:03}.

A HT condensation process of carbon nanoparticles process can be expected in supernovae \citep{Clayton:01}
or in the hot circumstellar environments of carbon-rich stars, such as Wolf-Rayet stars \citep{Cherchneff:00}.

In contrast, LT condensation realized by using cw laser-driven pyrolysis of hydrocarbons
favors a formation process with PAHs as precursors and particle-forming elements, which condense on the surface of
carbonaceous seeds via physi- and chemisorption. As a result, large carbonaceous
grains are formed revealing well developed planar and slightly bent graphene layers in their interior.
The samples produced in the present study are very valuable for a comparison with astrophysical data.
Particularly noteworthy is the fact that LT laser pyrolysis experiments can produce a wealth of PAHs.
The experimental conditions are comparable to those encountered in circumstellar
environments around evolved stars (AGB stars) suggesting that, in these environments, the soot formation is governed by the synthesis of PAHs and their condensation on the surface of larger grains.

In circumstellar shells of AGB stars, the pressure is assumed to be
lower than in the laboratory studies. However, our total pressure is
not comparable to the total pressures in the atmospheres of
dust-forming shells since we use a quenching gas to concentrate the
condensation within a relatively small volume. The lower pressure in
circumstellar condensation zones provokes the condensation to take
place within longer distances and timescales compared to laboratory
experiments, however, it is not assumed that this results in a
change of the  condensate structure. HT condensation experiments
performed in different pressures ranges (at $\sim$3 and
$\sim$300\,mbar) do not show a difference in the structure of the
condensing species and, therefore, they support this assumption.
Furthermore, we have to keep in mind that the graphite condensation
is pressure-independent and much more sensitive to the C/O ratio in
the dust-forming shells \citep{Lodders:99}. Furthermore, the
condensation parameters and, in particular, the pressure as well as
the degree of dust condensation can strongly vary in the
dust-forming zone due to pulsations of the stellar interior
\citep{Nowotny:05b}.

Low-temperature condensation is a very likely formation process of
soot and PAHs in AGB stars. Condensation temperatures in our
laboratory studies were found to be very similar to the temperature
range for carbon dust condensation in carbon-rich AGB stars,
predicted by \citet{Cherchneff:99} to start at temperatures of
1700\,K.

\acknowledgements This work was supported by a cooperation between
the Max-Planck-Institut f\"ur Astronomie and the FSU Jena as well as
by the Deutsche For\-schungs\-ge\-mein\-schaft (Hu 474/21-1).
Furthermore, we would like to thank Dr. H.J. R\"ader,
Max-Planck-Institut f\"ur Polymerchemie, Mainz, for the MALDI-TOF
measurements.

\end{document}